\documentclass[aps,twocolumn,prc,showpacs]{revtex4}
\usepackage{mathrsfs}
\usepackage{longtable,lscape}
\usepackage{txfonts}
\usepackage{amssymb}
\usepackage{indentfirst}
\usepackage{graphicx,,booktabs}
\usepackage{color}
\usepackage{amssymb}
\usepackage{epsfig}

\newcommand{\vsig}{\mbox{\boldmath$\sigma$\unboldmath}}

\newcommand{\be}{\begin{equation}}
\newcommand{\ee}{\end{equation}}
\newcommand{\bea}{\begin{eqnarray}}
\newcommand{\eea}{\end{eqnarray}}
\newcommand{\bean}{\begin{eqnarray*}}
\newcommand{\eean}{\end{eqnarray*}}

\newcommand{\gapproxeq}{\lower
.7ex\hbox{$\;\stackrel{\textstyle >}{\sim}\;$}}
\newcommand{\lapproxeq}{\lower
.7ex\hbox{$\;\stackrel{\textstyle <}{\sim}\;$}}

\begin{document}
\title{Combined analysis of the $\pi^-p\rightarrow K^0\Lambda$, $\eta n$ reactions in a chiral quark model}

\author{
Li-Ye Xiao, Fan Ouyang and Xian-Hui Zhong \footnote {E-mail:
zhongxh@hunnu.edu.cn}}  \affiliation{ 1) Department
of Physics, Hunan Normal University,  Changsha 410081, China }

\affiliation{ 2) Synergetic Innovation
Center for Quantum Effects and Applications (SICQEA), Changsha 410081,China}

\affiliation{ 3) Key Laboratory of
Low-Dimensional Quantum Structures and Quantum Control of Ministry
of Education, Changsha 410081, China}

\begin{abstract}

A combined analysis of the reactions $\pi^-p\rightarrow K^0\Lambda$ and
$\eta n$ is carried out with a chiral quark model. The data in the center-of-mass
(c.m.) energy range from threshold up to $W\simeq 1.8$ GeV
are reasonably described. For $\pi^-p\rightarrow K^0\Lambda$,
it is found that $N(1535)S_{11}$ and $N(1650)S_{11}$ paly crucial roles near threshold. The $N(1650)S_{11}$ resonance
contributes to the reaction through configuration mixing with $N(1535)S_{11}$.
The constructive interference between $N(1535)S_{11}$ and $N(1650)S_{11}$
is responsible for the peak structure around threshold in the total cross section.
The $n$-pole, $u$- and $t$-channel backgrounds provide significant contributions to the reaction as well.
While, for the $\pi^-p\rightarrow \eta n$ process, the ``first peak"
in the total cross section is dominant by $N(1535)S_{11}$, which has a
sizeable destructive interference with $N(1650)S_{11}$. Around $P_\pi \simeq $
1.0 GeV/c ($W\simeq 1.7$ GeV), there seems to be
a small bump structure in the total cross section,  which might be explained by the
interference between the $u$-channel and $N(1650)S_{11}$. The $N(1520)D_{13}$ resonance
notably affects the angle distributions of the cross sections, although less
effects are seen in the total cross section. The role of $P$-wave state $N(1720)P_{13}$ should be further confirmed
by future experiments. If $N(1720)P_{13}$ has a narrow width of $\Gamma\simeq 120$ MeV
as found in our previous work by a study of the $\pi^0$ photoproduction processes, obvious
evidence should be seen in the $\pi^-p\rightarrow K^0\Lambda$ and
$\eta n$ processes as well.

\end{abstract}


 \maketitle

\section{Introduction}{\label{introduction}}

Understanding of the baryon spectrum and searching for the missing
nucleon resonances and new exotic states are favored topics in
hadronic physics~\cite{Klempt:2009pi}. Pion-nucleon ($\pi N$)
scattering provides us an important tool to study the light baryon
spectrum. Most of our current knowledge about
the nucleon resonances listed in the Review of Particle Physics by
the Particle Data Group (PDG)~\cite{Agashe:2014kda} was extracted from partial wave
analyses of the $\pi N$ scattering. As we know, in the constituent quark model a
rich spectrum of nucleon resonances is predicted~\cite{Isgur:1978xj,Isgur:1977ef,Capstick:1986bm},
however, some of them are still missing. To look for the missing resonances and
deal with the unresolved issues in hadronic physics, more precise
measurements of pion-induced reactions are suggested recently in
Ref.~\cite{Briscoe:2015qia}, meanwhile, reliable partial wave
analyses are the same important as the precise measurements for us
to extract the information on resonance properties.

In the $\pi N$ scattering, the $\pi^-p\to K^0\Lambda$ and $\eta n$
reactions are of special interest in hadronic physics. The reasons are as follows.
i) Only the baryon resonances with isospin $I=1/2$ contribute to these two reactions due to the isospin
selection rule. Thus, the $\pi^-p\to K^0\Lambda$ and $\eta n$
processes provide us a rather clear place to extract the properties of nucleon resonances
without interferences from the $\Delta^*$ states. ii) The
$\pi^-p\to K^0\Lambda$ and $\eta n$ reactions can let us obtain
information on strong couplings of nucleon resonances to $K\Lambda$
and $\eta N$ channels, which can not be obtained in the elastic $\pi N$
scattering. iii) Furthermore, in these reactions
one may find evidence of some resonances which couple only
weakly to the $\pi N$ channel. Hence, many studies about these
two reactions have been carried out.

In experiments, some measurements of $\pi^-p\to K^0\Lambda$ and
$\eta n$ were carried out about 20 or 30 years ago, the
data had been collected in~\cite{Ronchen:2012eg}. For the $\pi^-p\to
K^0\Lambda$ reaction, the data of total cross section and
differential cross sections can be obtained in the whole resonance
range~\cite{Bertanza:1962pt,Jones:1971zm,Knasel:1975rr,Baker:1978qm,Saxon:1979xu,Binford:1969ts,B. Nelson,A. Baldini},
furthermore, some data of $\Lambda$ polarization are also obtained
in the energy region $W< 1.8$ GeV~\cite{Baker:1978qm}. While for the
$\pi^-p\to \eta n$ reaction, a few precise data on the differential
cross sections and total cross section can be obtained near the
$N(1535)S_{11}$ mass threshold from Crystal Ball spectrometer at
BNL~\cite{Prakhov:2005qb}, however, other old data~
\cite{Brown:1979ii,Deinet:1969cd,Feltesse:1975nz,Debenham:1975bj,Bulos:1970zk,
Richards:1970cy} might be
problematic over the whole energy range for its limited angle
coverage and uncontrollable uncertainties~\cite{Clajus:1992dh}. Thus, to get reliable
information on the resonance properties, a combined analysis of the
$\pi^-p\to K^0\Lambda$ and $\eta n$ reactions is necessary before
new precise data are obtained.

In theory, various theoretical approaches have been applied to
analyze the $\pi^-p\to K^0\Lambda$ and/or $\eta n$ reactions, such
as the chiral unitary model~\cite{Hyodo:2003qa}, $K$-matrix
methods~\cite{Sotona:1988fm,Feuster:1997pq,Shrestha:2012va,
Shrestha:2012ep,Penner:2002ma,Shklyar:2005xg,Shklyar:2006xw,
Shklyar:2004dy,Arndt:2003if,Arndt:2005dg,Shklyar:2012js}, dynamical
coupled-channel models~\cite{Gasparyan:2003fp,
Chiang:2004ye,JuliaDiaz:2006is, Ronchen:2012eg,Durand:2008es,
Vrana:1999nt,Durand:2008kw}, the
BnGa approach~\cite{Anisovich:2013tij}, chiral quark
model~\cite{Zhong:2007fx,He:2008uf}, and other effective
approachs~\cite{Wu:2014yca,Lu:2013jva,Ceci:2005vf}, etc. However, the analyses
from different models claim a different resonance content in the
reactions~\cite{Briscoe:2015qia}. For example, in $\pi^-p\to K^0\Lambda$ reaction the large
peak near $W=1.7$ GeV is explained as the dominant contributions
from both $N(1650)S_{11}$ and $N(1710)P_{11}$ in
Ref.~\cite{Shrestha:2012va}; from $N(1720)P_{13}$ in
Ref.~\cite{Penner:2002ma}; however, from $N(1535)S_{11}$,
$N(1650)S_{11}$ and $N(1720)P_{13}$ in Ref.~\cite{Wu:2014yca}.
While in $\pi^-p\to \eta n$ reaction, the second bump structure near
$W=1.7$ GeV is explained as the dominant contributions of
$N(1710)P_{11}$ in Refs.~\cite{Penner:2002ma,Ronchen:2012eg,
Shklyar:2012js}; in Ref.~\cite{He:2008uf} it is argued that this
structure is mainly caused by $N(1720)P_{13}$; however, in
Ref.~\cite{Durand:2008kw,Durand:2008es} it is predicted that this
structure is due to the interference between $N(1535)S_{11}$ and
$N(1650)S_{11}$. To clarify these puzzles in the reactions, more
theoretical studies are needed.

In this work, we carry a combined study of $\pi^-p\to
K^0\Lambda$ and $\eta n$ reactions within a chiral quark model.
Firstly, we hope to obtain a better understanding of the reaction mechanism
for these reactions. In our previous work~\cite{Zhong:2007fx},
we first extended the chiral quark model
to study of the $\pi^-p\to\eta n$ reaction, where we only obtained reasonable results
near threshold. In order to further
understand the reaction at higher resonance region and get more solid predictions, we
need revisit this reaction by combining with the reaction $\pi^-p\rightarrow K^0\Lambda$.
Secondly, we expect to further confirm the extracted properties of $N(1535)S_{11}$ from the $\eta$-
and $\pi$-meson photoproduction processes in our previous works~\cite{Zhong:2011ti,Xiao:2015gra}.
Where we predicted that $N(1535)S_{11}$ might be explained as a
mixing three-qaurk state between representations of $[70,^28]$ and
$[70,^48]$ in the quark model. However, in the literature it is
argued that it may contain a large admixture of pentaquark component
for its large couplings to $\eta N$ and $K\Lambda$
channels~\cite{Liu:2005pm,Zou:2006tw}. The
$N(1535)S_{11}$ resonance is also considered as a dynamically
generated state in the chiral unitary models
~\cite{Inoue:2001ip,Nieves:2001wt,Doring:2009uc,Geng:2008cv}. Finally, we hope
to extract some information on $N(1720)P_{13}$ in the $\pi^-p\to
K^0\Lambda$ and $\eta n$ reactions and test its properties obtained
by us in the $\pi$-meson photoproduction processes, where we found
the $N(1720)P_{13}$ resonance might favor a narrow width of
$\Gamma\simeq 120$ MeV~\cite{Xiao:2015gra}, which about a factor of 2 narrower than the
world average value from PDG~\cite{Agashe:2014kda}.

In the chiral quark model, an effective chiral Lagrangian is
introduced to account for the quark-pseudoscalar-meson coupling.
Since the quark-meson coupling is invariant under the chiral
transformation, some of the low-energy properties of QCD are
retained. There are several outstanding features for this model~\cite{Zhong:2007fx}.
One is that only a very limited number of adjustable parameters will appear in this
framework. In particular, only one overall parameter is needed for the nucleon
resonances to be coupled to the pseudoscalar mesons in the SU(6)$\otimes$O(3)
symmetry limit. This distinguishes from hadronic models where each resonance
requires one additional coupling constant as free parameter.
Another one is that all the nucleon resonances can be treated consistently
in the quark model. Thus, it has predictive powers when
exposed to experimental data, and information about the
resonance structures can be extracted. The chiral quark model has been well developed and widely
applied to meson photoproduction reactions
~\cite{Li:1994cy,Li:1995si,Li:1995vi,Li:1997gd,Zhao:2002id,Zhao:1998fn,Zhao:2001jw,Zhao:1998rt,
Li:1998ni,Saghai:2001yd,Zhao:2000iz,He:2008ty,Zhong:2011ti,Zhong:2011ht,Xiao:2015gra}. Recently, this model
has been successfully extended to $\pi N$ and $KN$ reactions as
well~\cite{Zhong:2007fx,Zhong:2008km,Zhong:2013oqa,Xiao:2013hca}.

This work is organized as follows. The model is
reviewed in Sec.II. Then, in Sec.III, our numerical results and
analyses are presented and discussed. Finally, a summary is
given in Sec.IV.

\begin{table}
\begin{center}
\caption{ \label{gfactor} $g$ factors extracted in the symmetric quark model.}
\begin{tabular}{ccc}
\hline\hline
Factor ~~~& \underline{$\pi^-p\rightarrow K^0\Lambda$} ~~~&\underline{$\pi^-p\rightarrow \eta n$} \\
$g_{s1}$   &$\frac{\sqrt{6}}{2}$  &$1.0$\\
$g_{s2}$   &$\frac{\sqrt{6}}{3}$  &$\frac{2}{3}$\\
$g_{\nu1}$ &$\frac{\sqrt{6}}{2}$  &$\frac{5}{3}$\\
$g_{\nu2}$ &$\frac{\sqrt{6}}{3}$  &$0.0$\\
$g_{s1}^u$ & 0.0                    &$1.0$\\
$g_{s2}^u$ &$\frac{\sqrt{6}}{3}$  &$\frac{2}{3}$\\
$g_{\nu1}^u$& 0.0                   &$\frac{5}{3}$\\
$g_{\nu2}^u$&$-\frac{\sqrt{6}}{3}$&$0.0$\\
$g_t^s$ &$\frac{\sqrt{6}}{2}$     &$1.0$\\
$g_t^v$ &$\frac{\sqrt{6}}{2}$     &$\frac{5}{3}$\\
\hline\hline
\end{tabular}
\end{center}
\end{table}

\begin{table*}
\begin{center}
\caption{ \label{amplitudes} The $s$-channel resonance amplitudes within $n$=$2$ shell for the $\pi^-p\rightarrow K^0
\Lambda,\eta n$ processes.
We have defined $M_{S}\equiv\left[\frac{\omega_i}{\mu_q}-|\mathbf{A}_{\mathrm{in}}|\frac{2|\mathbf{k}|}{3\alpha^2}\right]
\left[\frac{\omega_f}{\mu_q}-|\mathbf{A}_{\mathrm{out}}|
\frac{2|\mathbf{q}|}{3\alpha^2}\right]$, $M_P=
M_D\equiv\frac{|\mathbf{A}_{\mathrm{out}}||\mathbf{A}_{\mathrm{in}}|}{|\mathbf{k}||\mathbf{q}|}$,
$M_{P0}\equiv\left[\frac{\omega_i}{\mu_q}-|\mathbf{A}_{\mathrm{in}}|\frac{|\mathbf{k}|}{\alpha^2}\right]
\left[\frac{\omega_f}{\mu_q}-|\mathbf{A}_{\mathrm{out}}|
\frac{|\mathbf{q}|}{\alpha^2}\right]$,
$M_{P2}\equiv\left[\frac{\omega_i}{\mu_q}-|\mathbf{A}_{\mathrm{in}}|\frac{2|\mathbf{k}|}{5\alpha^2}\right]
\left[\frac{\omega_f}{\mu_q}-|\mathbf{A}_{\mathrm{out}}|
\frac{2|\mathbf{q}|}{5\alpha^2}\right]$,
$M_F\equiv|\mathbf{A}_{\mathrm{out}}||\mathbf{A}_{\mathrm{in}}|\frac{|\mathbf{k}||\mathbf{q}|}{\alpha^4}$
, $P'_{l}(z)\equiv\frac{\partial P_l(z)}{\partial z}$,
$X_{S1}\equiv\left[\cos\theta_S+\sin\theta_S\right]\left[2\cos\theta_S-\sin\theta_S\right]$,
$X_{S2}\equiv\left[\sin\theta_S-\cos\theta_S\right]\left[2\sin\theta_S+\cos\theta_S\right]$.
The functions $\mathbf{A}_{\mathrm{in}}$ and $\mathbf{A}_{\mathrm{out}}$ have been defined in~\cite{Zhong:2013oqa}. The $\mu_q$ is a reduced
mass at the quark level, which equals to $1/\mu_q=1/m_u+1/m_s$ for $K$ productions,
while $1/\mu_q=2/m_u$ for $\eta$ productions. $P_l(z)$ is the Legendre function with $z=\cos\theta$.
}
\begin{tabular}{c|c|c|c|c}
\hline\hline
$[N_6, ^{2S+1}N_3, n, l]$ &$l_{2I,2J}$ &$\mathcal{O}_R$ &$\pi^-p\rightarrow K^0 \Lambda$ &$\pi^-p\rightarrow \eta n$\\
\hline
$|70,^28,0,0\rangle$ &$P_{11}(n)$ &$f(\theta)$ &$\frac{5}{\sqrt{6}}M_P|\mathbf{k}||\mathbf{q}|P_1(z)$ &$\frac{5}{3}M_P|\mathbf{k}||\mathbf{q}|P_1(z)$\\
$$ &$$ &$g(\theta)$ &$\frac{5}{\sqrt{6}}M_P|\mathbf{k}||\mathbf{q}|\sin\theta P'_1(z)$ &$\frac{5}{3}M_P|\mathbf{k}||\mathbf{q}|\sin\theta P'_1(z)$\\
\hline
$+\cos\theta_S |70,^28,1,1\rangle$ &$N(1535)S_{11}$ &$f(\theta)$ &$\frac{\sqrt{6}}{12}\cos^2\theta_S M_S\alpha^2$ &$\frac{1}{6}X_{S1}M_S\alpha^2$\\
$-\sin\theta_S |70,^48,1,1\rangle$  &$$ &$g(\theta)$ &$-$ &$-$\\
$+\cos\theta_S|70,^48,1,1\rangle$ &$N(1650)S_{11}$ &$f(\theta)$ &$\frac{\sqrt{6}}{12}\sin^2\theta_S M_S\alpha^2$ &$\frac{1}{6}X_{S2}M_S\alpha^2$\\
$+\sin\theta_S |70,^28,1,1\rangle$  &$$ &$g(\theta)$ &$-$ &$-$\\
\cline{2-5}
$|70,^28,1,1\rangle$ &$N(1520)D_{13}$ &$f(\theta)$ &$\frac{20\sqrt{6}}{27}M_D\frac{|\mathbf{k}||\mathbf{q}|}{\alpha^2}P_2(z)$ &$\frac{76}{135} M_D\frac{|\mathbf{k}||\mathbf{q}|}{\alpha^2}P_2(z)$\\
$$  &$$ &$g(\theta)$ &$\frac{10\sqrt{6}}{27}M_D \frac{|\mathbf{k}||\mathbf{q}|}{\alpha^2} \sin(\theta)P'_2(z)$ &$\frac{38}{135} M_D\frac{|\mathbf{k}||\mathbf{q}|}{\alpha^2}\sin\theta P'_2(z)$\\
$|70,^48,1,1\rangle$ &$N(1700)D_{13}$ &$f(\theta)$ &-- &$-\frac{38}{1350}M_D\frac{|\mathbf{k}||\mathbf{q}|}{\alpha^2}P_2(z)$\\
$$  &$$ &$g(\theta)$ & -- &$-\frac{19}{1350}M_D\frac{|\mathbf{k}||\mathbf{q}|}{\alpha^2} \sin\theta P'_2(z)$\\
\cline{2-5}
$$ &$N(1675)D_{15}$ &$f(\theta)$ &$-$ &$-\frac{2}{15}M_D\frac{|\mathbf{k}||\mathbf{q}|}{\alpha^2}P_2(z)$\\
$$  &$$ &$g(\theta)$ &$-$ &$\frac{2}{45}M_D\frac{|\mathbf{k}||\mathbf{q}|}{\alpha^2}\sin\theta P'_2(z)$\\
\hline
$|56,^28,2,0\rangle$ &$N(1440)P_{11}$ &$f(\theta)$ &$\frac{5\sqrt{6}}{24\times27}M_{P0}|\mathbf{k}||\mathbf{q}|P_1(z)$ &$\frac{5}{12\times27}M_{P0}|\mathbf{k}||\mathbf{q}|P_1(z)$\\
$$  &$$ &$g(\theta)$ &$\frac{5\sqrt{6}}{24\times27}M_{P0}|\mathbf{k}||\mathbf{q}|\sin\theta P'_1(z)$ &$\frac{5}{12\times27}M_{P0}|\mathbf{k}||\mathbf{q}|sin\theta P'_1(z)$\\
$|70,^28,2,0\rangle$ &$N(1710)P_{11}$ &$f(\theta)$ &$\frac{2\sqrt{6}}{24\times27}M_{P0}|\mathbf{k}||\mathbf{q}|P_1(z)$ &$\frac{1}{3\times27}M_{P0}|\mathbf{k}||\mathbf{q}|P_1(z)$\\
$$  &$$ &$g(\theta)$ &$\frac{2\sqrt{6}}{24\times27}M_{P0}|\mathbf{k}||\mathbf{q}|\sin\theta P'_1(z)$ &$\frac{1}{3\times27}M_{P0}|\mathbf{k}||\mathbf{q}|\sin\theta P'_1(z)$\\
$|70,^48,2,2\rangle$ &$N(1880)P_{11}$ &$f(\theta)$ &$-$ &$-\frac{5}{4\times81}M_{P2}|\mathbf{k}||\mathbf{q}|P_1(z)$\\
$$  &$$ &$g(\theta)$ &$-$ &$-\frac{5}{4\times81}M_{P2}|\mathbf{k}||\mathbf{q}|\sin\theta P'_1(z)$\\
\cline{2-5}
$|70,^48,2,0\rangle$ &$N(?)P_{13}$ &$f(\theta)$ &$-$ &$-\frac{1}{2\times81}M_{P0}|\mathbf{k}||\mathbf{q}| P_1(z)$\\
$$  &$$ &$g(\theta)$ &$-$ &$\frac{1}{4\times81}M_{P0}|\mathbf{k}||\mathbf{q}|\sin\theta P'_1(z)$\\
$|56,^28,2,2\rangle$ &$N(1720)P_{13}$ &$f(\theta)$ &$\frac{25\sqrt{6}}{24\times27}M_{P2}|\mathbf{k}||\mathbf{q}|P_1(z)$ &$\frac{25}{4\times81}M_{P2}|\mathbf{k}||\mathbf{q}|P_1(z)$\\
$$  &$$ &$g(\theta)$ &$-\frac{25\sqrt{6}}{48\times27}M_{P2}|\mathbf{k}||\mathbf{q}|\sin\theta P'_1(z)$ &$-\frac{25}{8\times81}M_{P2}|\mathbf{k}||\mathbf{q}|sin\theta P'_1(z)$\\
$|70,^28,2,2\rangle$ &$N(1900)P_{13}$ &$f(\theta)$ &$\frac{10\sqrt{6}}{24\times27}M_{P2}|\mathbf{k}||\mathbf{q}|P_1(z)$ &$\frac{5}{81}M_{P2}|\mathbf{k}||\mathbf{q}|P_1(z)$\\
$$  &$$ &$g(\theta)$ &$-\frac{10\sqrt{6}}{48\times27}M_{P2}|\mathbf{k}||\mathbf{q}|\sin\theta P'_1(z)$ &$-\frac{5}{2\times81}M_{P2}|\mathbf{k}||\mathbf{q}|sin\theta P'_1(z)$\\
$|70,^48,2,2\rangle$ &$N(?)P_{13}$ &$f(\theta)$ &$-$ &$-\frac{5}{4\times81}M_{P2}|\mathbf{k}||\mathbf{q}|P_1(z)$\\
$$  &$$ &$g(\theta)$ &$-$ &$\frac{5}{8\times81}M_{P2}|\mathbf{k}||\mathbf{q}|\sin\theta P'_1(z)$\\
\cline{2-5}
$|56,^28,2,2\rangle$ &$N(1680)F_{15}$ &$f(\theta)$ &$\frac{3\sqrt{6}}{4}M_F|\mathbf{k}||\mathbf{q}|P_3(z)$ &$\frac{3}{2}M_F|\mathbf{k}||\mathbf{q}|P_3(z)$\\
$$  &$$ &$g(\theta)$  &$\frac{\sqrt{6}}{4}M_F|\mathbf{k}||\mathbf{q}|P'_3(z)$ &$\frac{1}{2}M_F|\mathbf{k}||\mathbf{q}|sin\theta P'_3(z)$\\
$|70,^28,2,2\rangle$ &$N(1860)F_{15}$ &$f(\theta)$ &$\frac{3\sqrt{6}}{10}M_F|\mathbf{k}||\mathbf{q}|P_3(z)$ &$\frac{6}{5}M_F|\mathbf{k}||\mathbf{q}|P_3(z)$\\
$$  &$$ &$g(\theta)$  &$\frac{\sqrt{6}}{10}M_F|\mathbf{k}||\mathbf{q}|\sin\theta P'_3(z)$ &$\frac{2}{5}M_F|\mathbf{k}||\mathbf{q}|\sin\theta P'_3(z)$\\
$|70,^48,2,2\rangle$ &$N(?)F_{15}$ &$f(\theta)$ &$-$ &$-\frac{3}{35}M_F|\mathbf{k}||\mathbf{q}|P_3(z)$\\
$$  &$$ &$g(\theta)$ &$-$ &$-\frac{1}{35}M_F|\mathbf{k}||\mathbf{q}|\sin\theta P'_3(z)$\\
\cline{2-5}
$|70,^48,2,2\rangle$ &$N(?)F_{17}$ &$f(\theta)$ &$-$ &$-\frac{18}{35}M_F|\mathbf{k}||\mathbf{q}|P_4(z)$\\
$$  &$$ &$g(\theta)$ &$-$ &$\frac{9}{70}M_F|\mathbf{k}||\mathbf{q}|\sin\theta P'_3(z)$\\
\hline\hline
\end{tabular}
\end{center}
\end{table*}

\section{Framework}

In this section, we give a brief review of the chiral quark model.
In this model, the $s$- and $u$-channel transition amplitudes can be
expressed as~\cite{Zhong:2007fx}
\begin{equation}\label{ms}
\mathbf{\cal M}_s=\sum_j\langle N_f|H_m^f|N_j\rangle\langle N_j|\frac{1}{E_i+\omega_i-E_j}H_m^i|N_i\rangle,
\end{equation}
\begin{equation}
\mathbf{\cal M}_u=\sum_j\langle N_f|H_m^i\frac{1}{E_i-\omega_i-E_j}|N_j\rangle\langle N_j|H_m^f|N_i\rangle,
\end{equation}
where $H_m^i$ and $H_m^f$ stand for the incoming and outgoing
meson-quark couplings, respectively. They might be described by the
effective chiral Lagrangian~\cite{Li:1997gd, Zhao:2002id},
\begin{equation}
H_m=\frac{1}{f_m}\bar{\psi}_j\gamma^j_{\mu}\gamma^j_5\psi_j\vec{\tau}\cdot\partial^{\mu}\vec{\phi}_m,
\end{equation}
where $\psi_j$ represents the $j$-th quark field in a hadron, $f_m$
is the meson's decay constant, and $\phi_m$ is the field of the
pseudoscalar-meson octet. The $\omega_i$ and $\omega_f$ are the
energies of the incoming and outgoing mesons, respectively.
$|N_i\rangle$, $|N_j\rangle$ and $|N_f\rangle$ stand for the
initial, intermediate, and final states, respectively, and their
corresponding energies are $E_i$, $E_j$, and $E_f$, which are the
eigenvalues of the nonrelativistic Hamiltonian of the constituent
quark model $\hat{H}$~\cite{Isgur:1978xj, Isgur:1977ef}. The $s$-
and $u$-channel transition amplitudes have been worked out in the
harmonic oscillator basis in Refs.~\cite{Zhong:2008km,
Zhong:2007fx,Zhong:2013oqa}, and the $g$ factors appearing in the $s$- and
$u$-channel amplitudes have been defined in Ref.~\cite{Zhong:2008km},
whose values are worked out and listed in Table~\ref{gfactor}.

The $t$-channel contributions of vector and/or scalar
exchanges are included in this work. The vector meson-quark and
scalar meson-quark couplings are give by
\begin{equation}
H_V=\bar{\psi}_j\left(a\gamma^{\nu}+\frac{b\sigma^{\nu\lambda}\partial_{\lambda}}{2m_q}\right)V_{\nu}\psi_j,
\end{equation}
\begin{equation}
H_S=g_{Sqq}\bar{\psi}_j\psi_jS,
\end{equation}
where $V$ and $S$ stand for the vector and scalar fields,
respectively. The constants $a$, $b$, and $g_{Sqq}$ are vector,
tensor, and scalar coupling constants, respectively. They are
treated as free parameters in this work.

Meanwhile, the $VPP$ and $SPP$ couplings ($P$ stands for a
pseudoscalar-meson) are adopted as
\begin{eqnarray}
H_{VPP}&=&-iG_VTr([\phi_m,\partial_\mu\phi_m]V^{\mu}),\\
H_{SPP}&=&\frac{g_{SPP}}{2m_\pi}\partial_\mu\phi_m\partial^\mu
\phi_m S,
\end{eqnarray}
where $G_V$ and $g_{SPP}$ are the $VPP$ and $SPP$ coupling
constants, respectively, to be determined by experimental data.

The $t$-channel transition amplitude has been given in Ref.~\cite{Zhong:2013oqa}.
In this work, the scalar $\kappa$-meson and vector $K^*$-meson exchanges are considered for
the $\pi^-p\rightarrow K^0\Lambda$ process, while the scalar
$a_0(980)$-meson exchange is considered for the $\pi^-p\rightarrow
\eta n$ process.

It should be pointed out that the amplitudes in terms of the harmonic
oscillator principle quantum number $n$ are sum of a set of SU(6)
multiplets with the same $n$. To obtain the contributions of
individual resonances, we need to separate out the
single-resonances-excitation amplitudes within each principle number
$n$ in the $s$ channel. Taking into account the width effects of the
resonances, the resonance transition amplitudes of $s$ channel can be
generally expressed as~\cite{Zhong:2013oqa, Zhong:2007fx}
\begin{equation}
\mathcal{M}_R^s=\frac{2M_R}{s-M_R^2+iM_R\Gamma_R}\mathcal{O}_Re^{-(\mathbf{k}^2+\mathbf{q}^2)/6\alpha^2},
\end{equation}
where $\sqrt{s}=E_i+\omega_i$ is the total energy of the system; $\mathbf{k}$ and
$\mathbf{q}$ stand for the momenta of incoming and outgoing
mesons, respectively; $\alpha$ is the harmonic oscillator strength; $\mathcal{O}_R$ is the
separated operators for individual resonances in the $s$ channel;
and $M_R$ is the mass of the $s$-channel resonance with a width
$\Gamma_R $. The transition amplitudes can be written in a standard
form~\cite{Hamilton:1963zz}:
\begin{equation}
\mathcal{O}_R=f(\theta)+i g(\theta)\vsig\cdot \mathbf{n},
\end{equation}
where $\vsig$ is the spin operator of the nucleon, while
$\mathbf{n}\equiv \mathbf{q}\times \mathbf{k}/|\mathbf{k}\times \mathbf{q}|$.
$f(\theta)$ and $g(\theta)$ stand for the non-spin-flip and spin-flip
amplitudes, respectively, which can be expanded in terms of the familiar
partial wave amplitudes $T_{l\pm}$ for the states with $J=l\pm 1/2$:
\begin{eqnarray}\label{f wave}
f(\theta)&=&\sum_{l=0}^\infty[(l+1)T_{l+}+lT_{l-}]P_l(\cos\theta),\\
g(\theta)&=&\sum_{l=0}^\infty[T_{l-}-T_{l+}]\sin\theta P_l^\prime(\cos\theta),
\end{eqnarray}
where $\theta$ is the scattering angle between $\mathbf{k}$ and
$\mathbf{q}$.

We have extracted the scattering amplitudes of the $s$-channel resonances within $n$=
$2$ shell for both $\pi^-p\to K^0 \Lambda$ and
$\pi^-p\rightarrow \eta n$, which have been listed in
Table~\ref{amplitudes}. It should be pointed out that
the contributions of $s$-channel resonances with a
$[70,^48]$ representation are forbidden in the
$\pi^- p \to K^0\Lambda$ reaction due to the Moorhouse selection
rule~\cite{Moorhouse:1966jn,Zhao:2006an}. Comparing these amplitudes of different
resonances with each other, one can easily find which states are the
main contributors to the reactions in the SU(6)$\otimes$O(3)
symmetry limit.

Finally, the differential cross section $d\sigma/d\Omega$ and
polarization of final baryon $P$ can be calculated by
\begin{eqnarray}\label{weifenjiemian}
\frac{d\sigma}{d\Omega}&=&\frac{(E_i+M_i)(E_f+M_f)}{64\pi^2s(2M_i)(2M_f)}\frac{|\mathbf{q}|}{|\mathbf{k}|}\frac{1}{2}
\times\sum_{\lambda_i,\lambda_f}\left|M_{\lambda_f,\lambda_i}\right|^2,\nonumber
\end{eqnarray}
\begin{eqnarray}\label{jihua}
P=2\frac{\mathrm{Im}[f(\theta)g^{\ast}(\theta)]}{|f(\theta)|^2+|g(\theta)|^2},
\end{eqnarray}
where $\lambda_i=\pm\frac{1}{2}$ and $\lambda_f=\pm\frac{1}{2}$ are
the helicities of the initial and final baryon states, respectively.

\begin{table}
\begin{center}
\caption{Resonance masses $M_R$ (MeV) and widths $\Gamma_R$ (MeV) in this work compared with
 the world average value from the PDG~\cite{Agashe:2014kda}.}\label{compare}
\begin{tabular}{ccccc}
\hline\hline
 resonance ~~~~~&$M_R$ ~~~~~&$\Gamma_R$ ~~~~~& $M_R$ (PDG)~~~~~& $\Gamma_R$ (PDG)   \\
\hline
$N(1535)S_{11}$~~~~~ &1524~~~~~ &$124$ ~~~~~ &$1535\pm10$ ~~~~~&$150\pm25$   \\
$N(1650)S_{11}$~~~~~ &1670~~~~~ &$119$ ~~~~~ &$1655_{-10}^{+15}$ ~~~~~&$140\pm30$   \\
$N(1520)D_{13}$~~~~~ &1515~~~~~ &$125$ ~~~~~ &$1515\pm5$  ~~~~~&$115^{+10}_{-15}$   \\
$N(1700)D_{13}$~~~~~ &1700~~~~~ &$150$ ~~~~~ &$1700\pm50$  ~~~~~&$150^{+100}_{-50}$   \\
$N(1675)D_{15}$~~~~~ &1685~~~~~ &$140$ ~~~~~ &$1675\pm5$  ~~~~~&$150^{+15}_{-20}$   \\
$N(1440)P_{11}$~~~~~ &1430~~~~~ &$350$ ~~~~~ &$1430\pm20$ ~~~~~&$350\pm100$   \\
$N(1710)P_{11}$~~~~~ &1710~~~~~ &$200$ ~~~~~ &$1710\pm30$ ~~~~~&$100^{+150}_{-50}$  \\
$N(1870)P_{11}$~~~~~ &1870~~~~~ &$235$ ~~~~~ &$1870\pm35$ ~~~~~&$235\pm65$  \\
$N(?)P_{13}$~~~~~ &2000~~~~~ &$200$ ~~~~~ &$?$ ~~~~~&$?$   \\
$N(1720)P_{13}$~~~~~ &1690~~~~~ &$400$ ~~~~~ &$1720^{+30}_{-20}$ ~~~~~&$250^{+150}_{-100}$   \\
$N(1900)P_{13}$~~~~~ &1900~~~~~ &$250$ ~~~~~ &$\sim1900$ ~~~~~&$\sim250$   \\
$N(?)P_{13}$~~~~~ &2040~~~~~ &$200$ ~~~~~ &$?$ ~~~~~&$?$   \\
$N(1680)F_{15}$~~~~~ &1680~~~~~ &$130$ ~~~~~ &$1685\pm5$ ~~~~~&$130\pm10$  \\
$N(1860)F_{15}$~~~~~ &1860~~~~~ &$270$ ~~~~~ &$1860^{+100}_{-40}$ ~~~~~&$270^{+140}_{-50}$   \\
$N(?)F_{15}$~~~~~ &2050~~~~~ &$200$ ~~~~~ &$?$ ~~~~~&$?$   \\
$N(?)F_{17}$~~~~~ &1990~~~~~ &$200$ ~~~~~ &$?$ ~~~~~&$?$   \\
\hline\hline
\end{tabular}
\end{center}
\end{table}

\begin{table}
\begin{center}
\caption{ Extracted partial decay width ratios for $N(1535)$ and $N(1650)$ resonances
compared with the values from the PDG~\cite{Agashe:2014kda}.}\label{pwr}
\begin{tabular}{c|cc cc}
\hline
 Resonance~~~&$\frac{\Gamma_{\eta N}}{\Gamma_{\pi N}}$
(ours)  ~~~&$\frac{\Gamma_{\eta N}}{\Gamma_{\pi N}}$ (PDG)
~~~&$\frac{\Gamma_{K\Lambda}}{\Gamma_{\pi N}}$ (ours)&$\frac{\Gamma_{K\Lambda}}{\Gamma_{\pi N}}$ (PDG)\\
\hline
$N(1535)S_{11}$  &1.57   ~~~&$1.20^{+0.29}_{-0.62}$  ~~~&/     ~~~&/   \\
$N(1650)S_{11}$  &0.23   ~~~&0.05-0.30               ~~~&0.20  ~~~&0.03-0.22    \\
\hline\hline
\end{tabular}
\end{center}
\end{table}

\begin{table*}
\begin{center}
\caption{ Reduced $\chi^2$ per data point of the full model and that with one resonance or one background switched off obtained in a global fit of the data of $\pi^-p\to K^0 \Lambda$ and $\pi^-p\to \eta n$. The corresponding partial $\chi^2s$ for the $\pi^-p\to K^0 \Lambda$ (labeled with $\chi^2_{K}$ )
and $\pi^-p\to \eta n$  (labeled with $\chi^2_{\eta}$ ) are also included.}\label{consequence}
\begin{tabular}{ccccccccc}
\hline\hline
 ~~~&full model    ~~~&$n-$pole ~~~&$N(1535)S_{11}$ ~~~&$N(1650)S_{11}$ ~~~&$N(1520)D_{13}$ ~~~&$N(1720)P_{13}$ ~~~&$u$-channel ~~~&$t$-channel\\
\hline
$\chi^2$        ~~~&5.98       ~~~&14.96 ~~~&265.03        ~~~&24.88     ~~~&14.53    ~~~&5.99  ~~~&15.79  ~~~&15.03   \\
$\chi^2_{\eta}$ ~~~&8.54       ~~~&13.36 ~~~&466.86        ~~~&40.31     ~~~&24.52    ~~~&8.07  ~~~&18.09  ~~~&10.49  \\
$\chi^2_{K}$    ~~~&2.78       ~~~&16.95  ~~~&12.75        ~~~&5.58      ~~~&2.05     ~~~&3.39  ~~~&12.92  ~~~&20.70  \\
\hline\hline
\end{tabular}
\end{center}
\end{table*}

\section{Calculation and analysis}

\subsection{Parameters}

In the calculation, the universal value of harmonic oscillator
parameter $\alpha$=0.4 GeV is adopted. The masses of the u, d, and s
constituent quarks are adopted as $m_u$=$m_d$=330 MeV and $m_s$=450 MeV,
respectively.

In our work, the $s$-channel resonance transition amplitude,
$\mathcal{O}_R$, is derived in the SU(6)$\otimes$O(3) symmetric
quark model limit. In reality, the symmetry of SU(6)$\otimes$O(3) is
generally broken due to, e.g.,  spin-dependent forces in the
quark-quark interaction. As a result, configuration mixing would
occur, which can produce an effect on our theoretical predictions.
According to our previous studies of the $\eta$ and $\pi^0$
photoproduction on the nucleons~\cite{Zhong:2011ti,Xiao:2015gra}, we found the
configuration mixings seem to be inevitable for the low-lying
$S$-wave nucleon resonances $N(1535)S_{11}$ and $N(1650)S_{11}$.
Thus, in this work we also consider configuration mixing effects in
the $S$-wave states and use the same mixing scheme as in our
previous works~\cite{Zhong:2011ti,Xiao:2015gra},
\begin{equation}\label{mixs}
\left(\begin{array}{c}S_{11}(1535)\cr  S_{11}(1650)
\end{array}\right)=\left(\begin{array}{cc} \cos\theta_S &-\sin\theta_S\cr \sin\theta_S & \cos\theta_S
\end{array}\right)
\left(\begin{array}{c} |70,^28,1/2^-\rangle \cr |70,^48,1/2^-\rangle
\end{array}\right),
\end{equation}
where $\theta_S$ is the mixing angle.
Then, the $s$-channel resonance transition amplitudes of the
$S$-wave states $N(1535)S_{11}$ and $N(1650)S_{11}$ are related to
the mixing angle $\theta_S$. These transition amplitudes have been
worked out and listed in Tab.~\ref{amplitudes}. The mixing angle
$\theta_S$ has been determined by fitting the data. The determined
value $\theta_S\simeq 26.9^\circ$ is consistent with that
suggested in our previous works~\cite{Zhong:2011ti,Xiao:2015gra}.

In the calculations, the quark-pseudoscalar-meson couplings are the
overall parameters in the $s$- and $u$-channel transitions. However,
they are not totally free ones. They can be related to the hadronic
couplings via the Goldberger-Treiman relation~\cite{Goldberger:1958tr}
\begin{eqnarray}
g_{mNN}=\frac{g^m_AM_N}{F_m},
\end{eqnarray}
where $m$ stands for the pseudoscalar mesons, $\eta$, $\pi$ and $K$;
$g^m_A$ is the axial vector coupling for the meson; and $F_m$ is the
meson decay constant, which can be related to $f_m$ defined earlier
by $F_m=f_m/\sqrt{2}$.

It should be pointed out that the $\pi NN$ coupling constant $g_{\pi NN}$ is
a well-determined number, $g_{\pi NN}=13.48$, thus, we fix it in
our calculations. While for the other two coupling
constants $g_{KN\Lambda}$ and $g_{\eta NN}$, there are larger
uncertainties. We determine them by fitting the data of the $\pi^- p
\to K^0\Lambda,\eta n$ processes, respectively. We get that
\begin{eqnarray}
g_{KN\Lambda}\simeq 6.87, ~~~~~~ g_{\eta NN}\simeq 2.50.
\end{eqnarray}
The coupling constant $g_{\eta NN}$ extracted in present work is
consistent with that extracted from the $\eta$ meson photoproduction
on nucleons in our previous work~\cite{Zhong:2011ti}, and also in
good agreement with the determinations in
Refs.~\cite{Li:1995vi,Tiator:1994et,Zhu:2000eh,Piekarewicz:1993ad}.
And the coupling constant $g_{KN\Lambda}$ extracted by us is
consistent with that extracted from the $K$ meson photoproduction on
nucleons in Refs.~\cite{Li:1995si,JuliaDiaz:2006is,Chiang:2001pw}.

In the $t$ channel of the $\pi^-p\to K^0\Lambda$ process, there are
two free parameters, $G_{V}a$ and $g_{SPP}g_{Sqq}$, which come from
$K^*$ and $\kappa$ exchanges, respectively. By fitting the data, we
get $G_{V}a\simeq 3.0$ and $g_{SPP}g_{Sqq}$$\simeq 36.4$, which are
close to our previous determinations in the $K^-p$ scattering~\cite{Zhong:2013oqa}. While, in
the $t$ channel of the $\pi^-p\to \eta n$ process, the parameter
from the $a_0$ exchange $g_{a_0\pi\eta}g_{a_0NN}$ are adopted a
commonly used value $g_{a_0\pi\eta}g_{a_0NN}\simeq$100 based on
Refs.~\cite{Krehl:1999km, Gasparyan:2003fp}.

In the $u$ channel, it is found that contributions from the $n\geq
1$ shell resonances are negligibly small and insensitive to their
masses. Thus, the $n=1$ and the $n=2$ shell resonances are treated
as degeneration. In the calculations, we take $M_1=1650$ MeV
($M_2=1750$ MeV) for the degenerate mass of $n=1$ ($n=2$) shell
resonances.

In the $s$ channel, the masses and widths of the nucleon resonances
are taken from the PDG~\cite{Agashe:2014kda}, or the constituent
quark model predictions~\cite{Capstick:1986bm} if no experimental
data are available. For the main resonances, we allow their masses
and widths to change in a proper range in order to better describe the data. The
determined values are listed in Tab. \ref{compare}. It is found that
the main resonance masses and widths extracted by us are in good
agreement with the world average values from
PDG~\cite{Agashe:2014kda}. One point should be emphasized that our
global fits of the $\pi^-p\to \eta n,K^0\Lambda$ reaction data seem
to favour a broad width $\Gamma\simeq 400$ MeV for $N(1720)P_{13}$,
however, this width is much broader than $\Gamma\simeq 120$ MeV extracted from the
neutral pion photoproduction processes in our previous
work~\cite{Xiao:2015gra}. A similar narrow width of $N(1720)P_{13}$
was also found by the CLAS Collaboration in the reaction $ep\to
ep' \pi^+\pi^-$~\cite{Ripani:2002ss}. We will further discuss whether a narrow width
state $N(1720)P_{13}$ is allowed or not in the $\pi^-p\to \eta
n,K^0\Lambda$ reactions later.

Combining the extracted coupling constants,
$g_{KN\Lambda}$ and $g_{\eta NN}$, and resonance masses, we further
determine some partial width ratios of the main contributors
$N(1535)S_{11}$ and $N(1650)S_{11}$ to the reactions, which have
been listed in Tab.~\ref{pwr}. Form the table we can see that the partial width ratios
$\Gamma_{\eta N}/\Gamma_{\pi N}$ and $\Gamma_{K\Lambda}/\Gamma_{\pi N}$ determined by us
are close to the upper limit of the values from the PDG
~\cite{Agashe:2014kda}.

\begin{figure*}[htbp]
\begin{center}
\centering \epsfxsize=16.8 cm \epsfbox{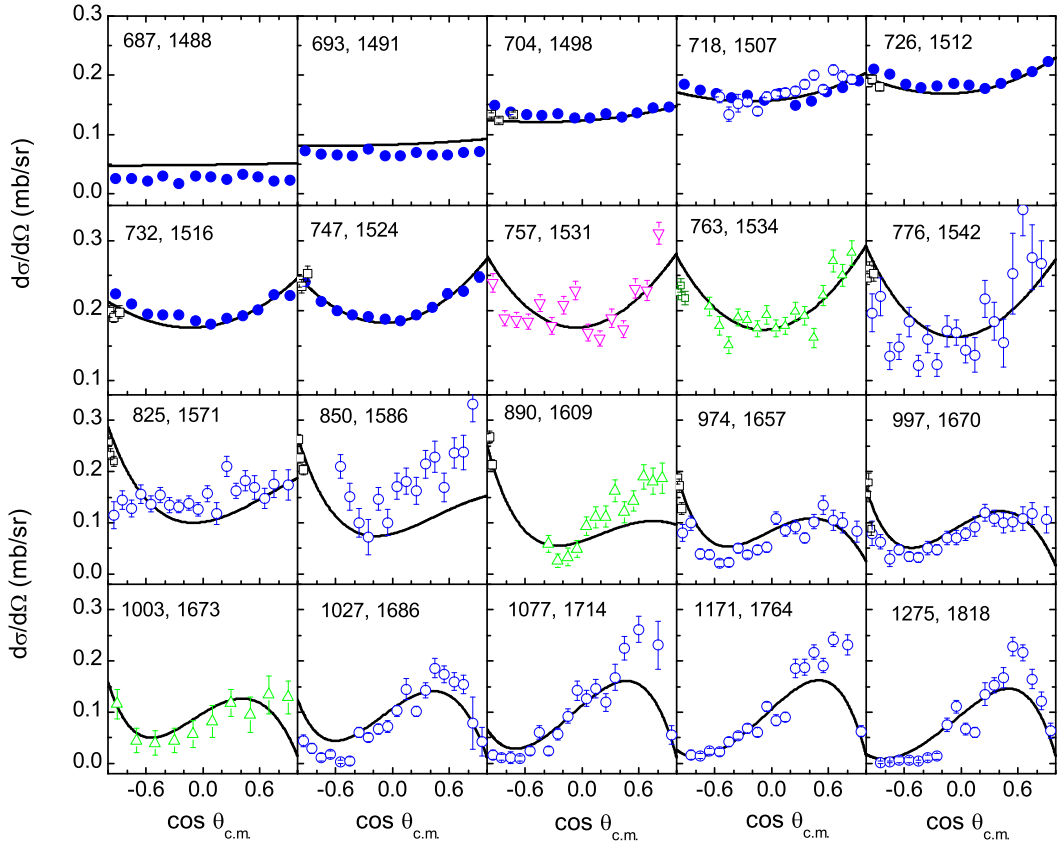} \vspace{-4.2cm}\caption{Differential cross sections of the
reaction $\pi^- p \to \eta n$ compared with the experimental data from Refs.~\cite{Brown:1979ii} (open
circle),~\cite{Deinet:1969cd} (open
up-triangles),~\cite{Feltesse:1975nz} (open down-triangles),
~\cite{Debenham:1975bj} (open squares) and the recent
experiment presented in Ref.~\cite{Prakhov:2005qb} (solid
circles). The first and second
numbers in each figure correspond to the $\pi^-$ beam momentum $P_\pi$
(MeV) and the $\pi N$ center-of-mass (c.m.) energy $W$ (MeV),
respectively. }  \label{diff-etan}
\end{center}
\end{figure*}

Finally, it should be pointed out that all adjustable parameters are
determined by globally fitting the measured differential cross
sections of the $\pi^-p\to \eta n, K^0\Lambda$ processes. All the
data sets used in our fits have been shown in Figs.~\ref{diff-KL-2}
and \ref{diff-etan}. The reduced $\chi^2$ per data point obtained in our fits has been
listed in Tab.~\ref{consequence}. To clearly see the role of one
component in the reactions, the $\chi^2s$ with one resonance or one background
switched off are also given in the table.

\begin{figure*}[ht]
\centering \epsfxsize=16.8 cm \epsfbox{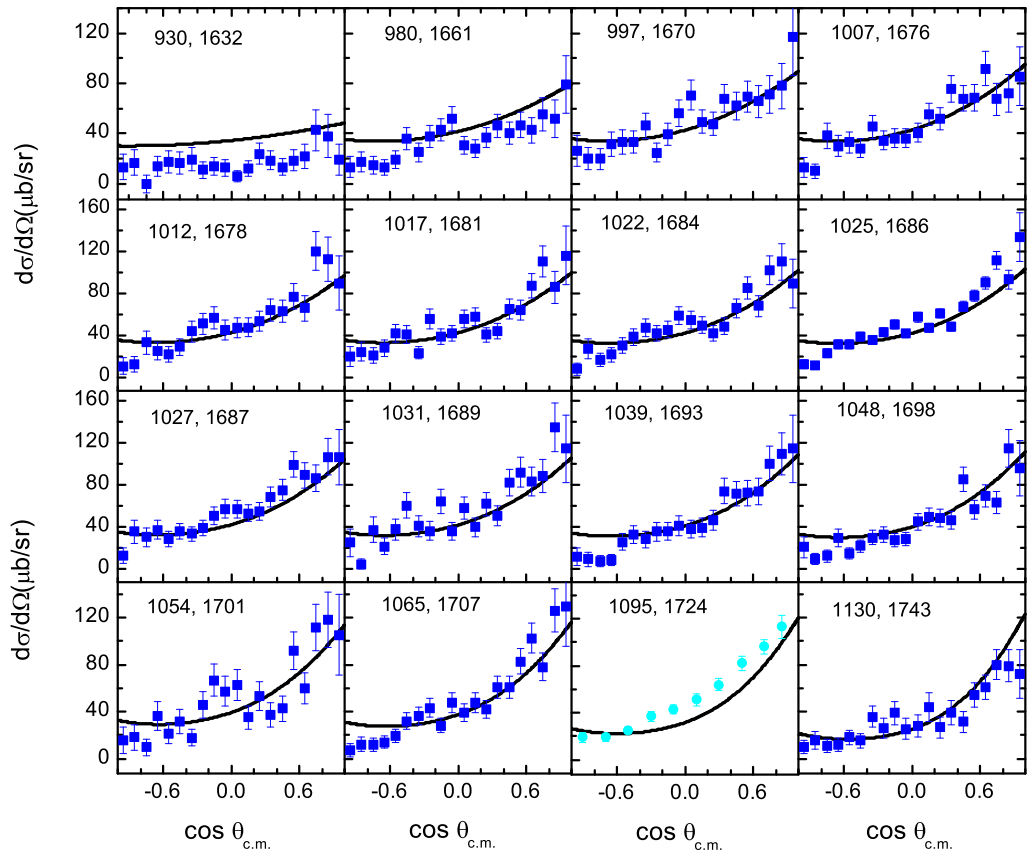}
\vspace{-1.2cm} \caption{Differential cross sections of the
reaction $\pi^- p \to K^0\Lambda$ compared with the experimental
data from Refs.~\cite{B. Nelson} (solid
squares) and \cite{Baker:1978qm} (solid circle). The first and second
numbers in each figure correspond to the $\pi^-$ beam momentum $P_\pi$
(MeV) and the $\pi N$ center-of-mass (c.m.) energy $W$ (MeV),
respectively.} \label{diff-KL}
\end{figure*}

\begin{figure}[htbp]
\includegraphics[scale=0.6]{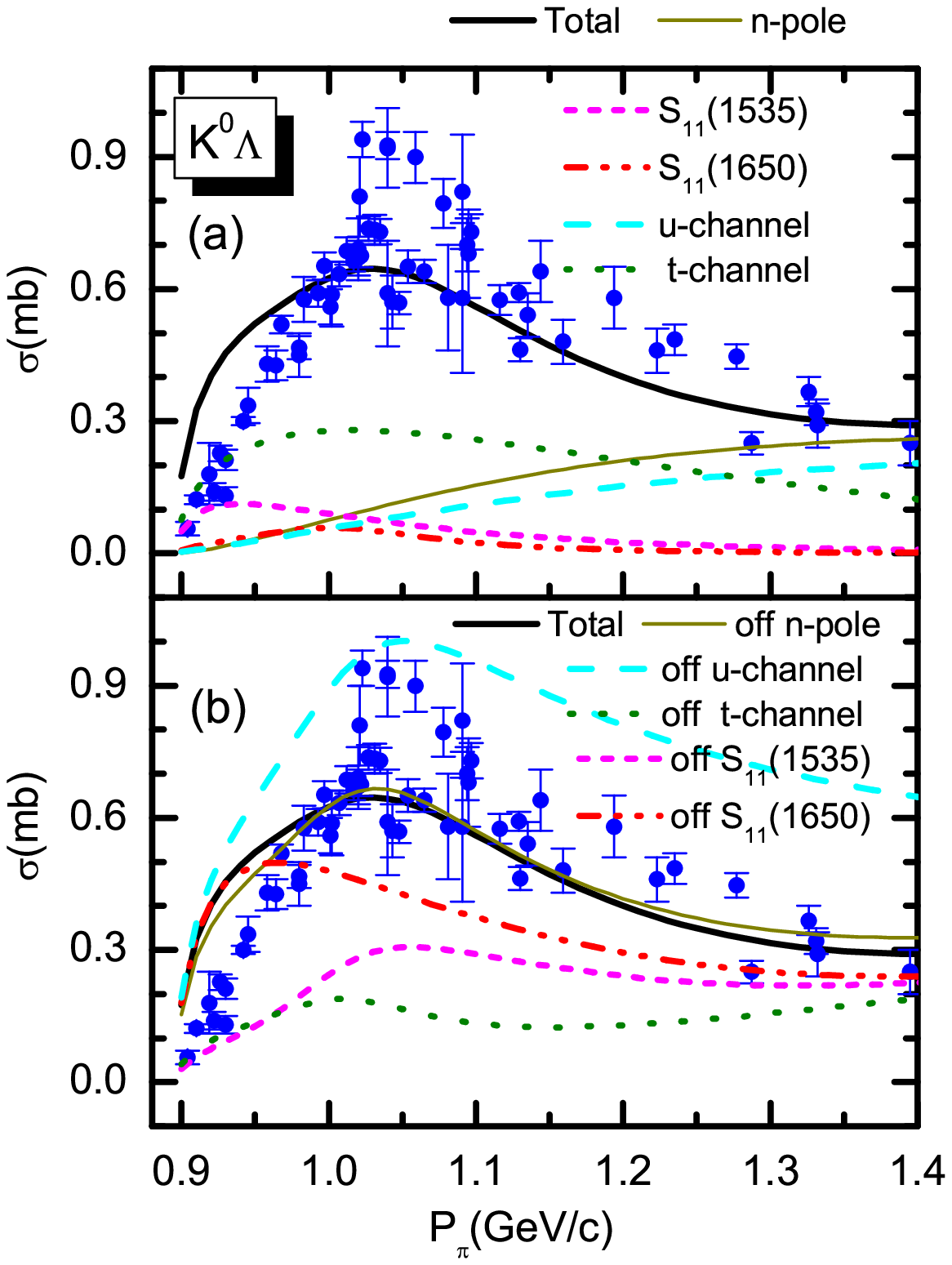}
\vspace{-0.4cm} \caption{ Total cross section of the reaction
$\pi^- p \to K^0\Lambda$ compared with experimental data from Ref.~\cite{A.
Baldini}. The bold solid curves correspond to the full model result.
In figure (a), exclusive cross sections for $S_{11}(1535)$,
$S_{11}(1650)$, nucleon pole, $u$ channel and $t$ channel are
indicated explicitly by the legends. In figure (b), the results by switching off the contributions of
$S_{11}(1535)$, $S_{11}(1650)$, nucleon pole, $u$ channel and $t$
channel are indicated explicitly by the legends.}
\label{total-kl}
\end{figure}

\subsection{$\pi^-p\rightarrow K^0\Lambda$}

The differential cross sections and total cross section of the
$\pi^- p \to K^0\Lambda$ process compared with experimental data are
shown in Figs.\ref{diff-KL} and \ref{total-kl}, respectively. From
these figures, it is found that the experimental data in the c.m.
energy range from threshold up to $W\simeq1.8$ GeV
are reasonably described within the chiral quark model.

\begin{figure}[htbp]
\begin{center}
\includegraphics[scale=0.6]{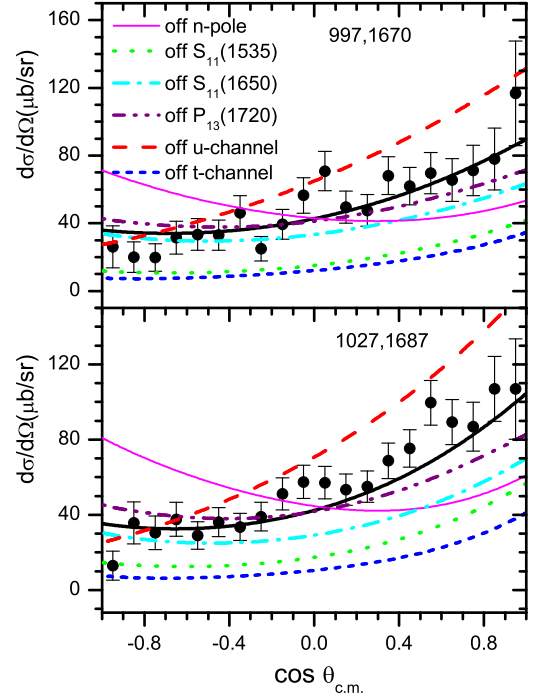}
\vspace{-0.8 cm} \caption{ Differential cross sections of the
reaction $\pi^- p \to K^0\Lambda$ compared with experimental
data~\cite{B. Nelson} at two energy points $P_{\pi}$ =997, 1027
MeV/c. The bold solid curves correspond to the full model result.
The predictions by switching off the contributions from
$N(1535)S_{11}$, $N(1650)S_{11}$, $N(1720)P_{13}$, and n-pole, $u$-
and $t$-channel backgrounds are indicated explicitly by the legends
in the figures.} \label{diff-KL-2}
\end{center}
\end{figure}

Obvious roles of the $S$-wave states $N(1535)S_{11}$ and
$N(1650)S_{11}$ are seen in the reaction. The constructive
interferences between $N(1535)S_{11}$ and $N(1650)S_{11}$ are
crucial to reproduce the bump structure near threshold in the total cross
section. Switching off the contributions of $N(1535)S_{11}$ or
$N(1650)S_{11}$, the cross sections around their mass thresholds are
notably underestimated (see Figs.~\ref{total-kl} and
\ref{diff-KL-2}). It should be pointed out that in the symmetric
quark model, the $N(1650)S_{11}$ resonance corresponds to the
$[70,^48]$ representation. In the SU(6)$\otimes$O(3) symmetry limit,
the contributions of $N(1650)S_{11}$ should be forbidden in the
$\pi^- p \to K^0\Lambda$ reaction due to the Moorhouse selection
rule~\cite{Moorhouse:1966jn,Zhao:2006an}. The obvious evidence of
$N(1650)S_{11}$ in the $\pi^- p \to K^0\Lambda$ reaction further
confirms that the SU(6)$\otimes$O(3) symmetry is broken, and the
configuration mixing between $N(1535)S_{11}$ and $N(1650)S_{11}$
should be necessary as suggested in our previous studies of
the meson photoproduction processes~\cite{Zhong:2011ti,Xiao:2015gra}.

\begin{figure}[htbp]
\begin{center}
\includegraphics[scale=0.7]{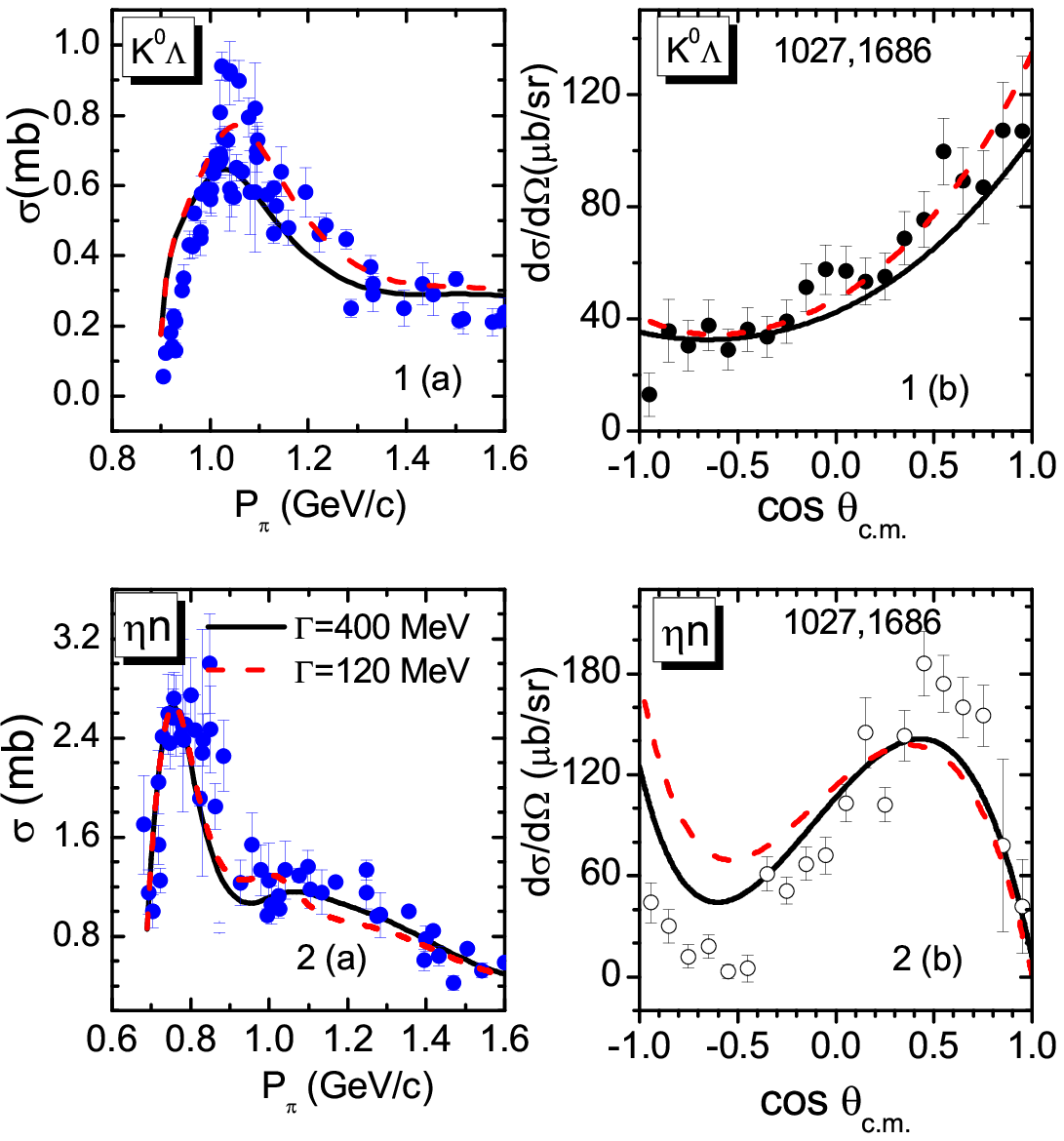}
\vspace{-0.5 cm} \caption{ Predictions for
the $\pi^- p\rightarrow K^0\Lambda$ and $\eta n$ reactions with
a narrow width $\Gamma=120$ MeV and a broad width $\Gamma=400$ MeV
for the resonance $N(1720)P_{13}$, respectively. The total cross sections
for $\pi^- p\rightarrow K^0\Lambda$ and $\eta n$ are plotted in
1(a) and 2 (a), respectively. While the differential cross sections at
$P_{\pi}$=1027 MeV/c for $\pi^- p\rightarrow K^0\Lambda$ and $\eta n$
are plotted in 1 (b) and 2 (b), respectively. The solid curves stand
for the results with a broad width $\Gamma=400$ MeV, while the dashed curves
for the results with a narrow width $\Gamma=120$ MeV.} \label{N(1720)}
\end{center}
\end{figure}

Furthermore, some contributions from $N(1720)P_{13}$ might be seen
in the differential cross sections. At backward angles, the
differential cross sections are slightly underestimated without its
contribution. For the large uncertainties of the data, here we can
not obtain solid information on $N(1720)P_{13}$. If we adopt a narrow
width $\Gamma\simeq 120$ MeV as suggested in our previous work~\cite{Xiao:2015gra}, from
Fig.~\ref{N(1720)} we see that the peak of the bump structure in the
total cross section becomes sharper, and around the mass threshold
of $N(1720)P_{13}$ the differential cross sections at forward angles are enhanced
significantly. Our theoretical predictions with a narrow width for
$N(1720)P_{13}$ are still consistent with the data within their
uncertainties. Thus, to finally determine the properties of
$N(1720)P_{13}$, we need more accurate measurements of the $\pi^- p
\to K^0\Lambda$ reaction.

The $n$-pole, $u$- and $t$-channel backgrounds play crucial roles in
the reaction as well. From the Figs.~\ref{total-kl} and
\ref{diff-KL-2}, one can see that switching off the $u$-channel
contribution, the cross sections should be strongly overestimated;
while switching off the $t$-channel contribution, the cross sections
will be underestimated draftily. The $n$-pole has obvious effects on
angle distributions of the cross sections in the whole energy region
what we considered. Without the $n$-pole contribution, the
differential cross sections at forward angles should be notably
underestimated, while those at backward angles should be
notably overestimated.

\begin{figure*}[htbp]
\begin{center}
\includegraphics[scale=1.2]{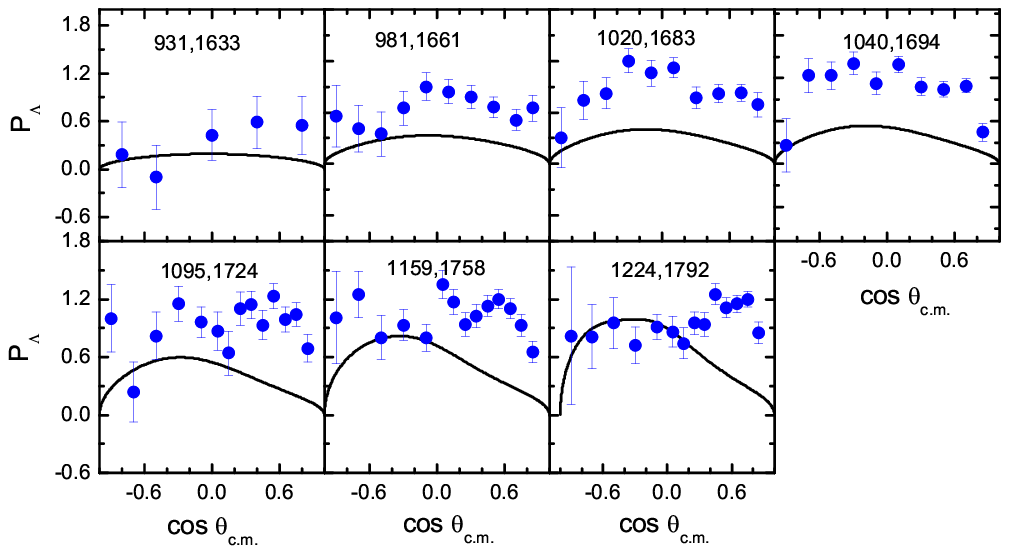}
\vspace{-0.2cm} \caption{$\Lambda$ polarization of the $\pi^- p
\to K^0\Lambda$ reaction compared with experimental
data~\cite{Baker:1978qm}. } \label{polarization}
\end{center}
\end{figure*}

In experiments, there are some old measurements for the $\Lambda$
polarization of the $\pi^- p \to K^0\Lambda$ reaction~\cite{Baker:1978qm}. In
Fig.~\ref{polarization} we compare our chiral quark model
predictions with the observations in the c.m. energy range $W< 1.8$ GeV.
We found that the experimental observations can be explained
reasonably. It should be emphasized that our theoretical
calculations seem to slightly underestimate the $\Lambda$ polarization.
This phenomenons also exist in the coupled-channel approach
calculations~\cite{Ronchen:2012eg}. Improved measurements and more
reliable experimental data of the polarization are needed to clarify
the discrepancies.

As a whole, obvious evidence of $N(1535)S_{11}$ and $N(1650)S_{11}$
are found in the $\pi^- p \to K^0\Lambda$ reaction, which is
consistent with the recent analysis within an
effective Lagrangian approach~\cite{Wu:2014yca}. It should be
emphasized that the $N(1650)S_{11}$ resonance contributes to the
reaction via the configuration mixing with
$N(1535)S_{11}$. The determined mixing angle is $\theta_S\simeq 26.9^\circ$.
Furthermore, remarkable contributions from the
backgrounds, $n$-pole, $u$- and $t$-channel, are found in the
reaction. There might be sizeable contributions from $N(1720)P_{13}$,
the present data can not determine whether it is a narrow or broad state.
No clear evidence from the other nucleon resonances, such as $N(1520)D_{13}$,
$N(1700)D_{13}$ and $N(1710)P_{11}$, is found in the reaction. Finally,
it should be mentioned that the $N(1710)P_{11}$ was considered as one
of the main contributors to $\pi^- p \to K^0\Lambda$ in the literature
~\cite{Ronchen:2012eg,Shrestha:2012va,Ceci:2005vf}, which is
in disagreement with our prediction and that from Refs.
~\cite{Penner:2002ma,Shklyar:2005xg,Wu:2014yca}.


\begin{figure}[htbp]
\begin{center}
\includegraphics[scale=0.55]{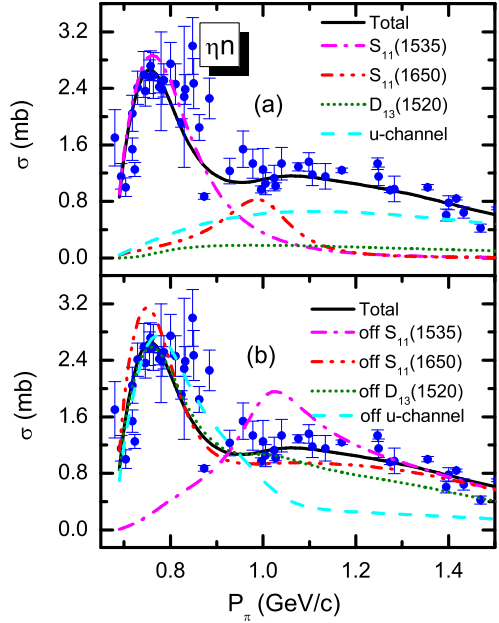}
\vspace{-0.2cm} \caption{Total cross section of the reaction $\pi^- p \to \eta n$ compared with experimental data~\cite{A. Baldini}. The bold solid curves correspond to the full model result. In figure (a), exclusive cross sections for $S_{11}(1535)$, $S_{11}(1650)$, $N(1520)D_{13}$, and $u$ channel are indicated explicitly by the legends. In figure (b), the results by switching off the contributions of $S_{11}(1535)$, $S_{11}(1650)$, $N(1520)D_{13}$, and $u$ channel are indicated explicitly by the legends. } \label{total-etan}
\end{center}
\end{figure}

\begin{figure}[htbp]
\begin{center}
\includegraphics[scale=0.55]{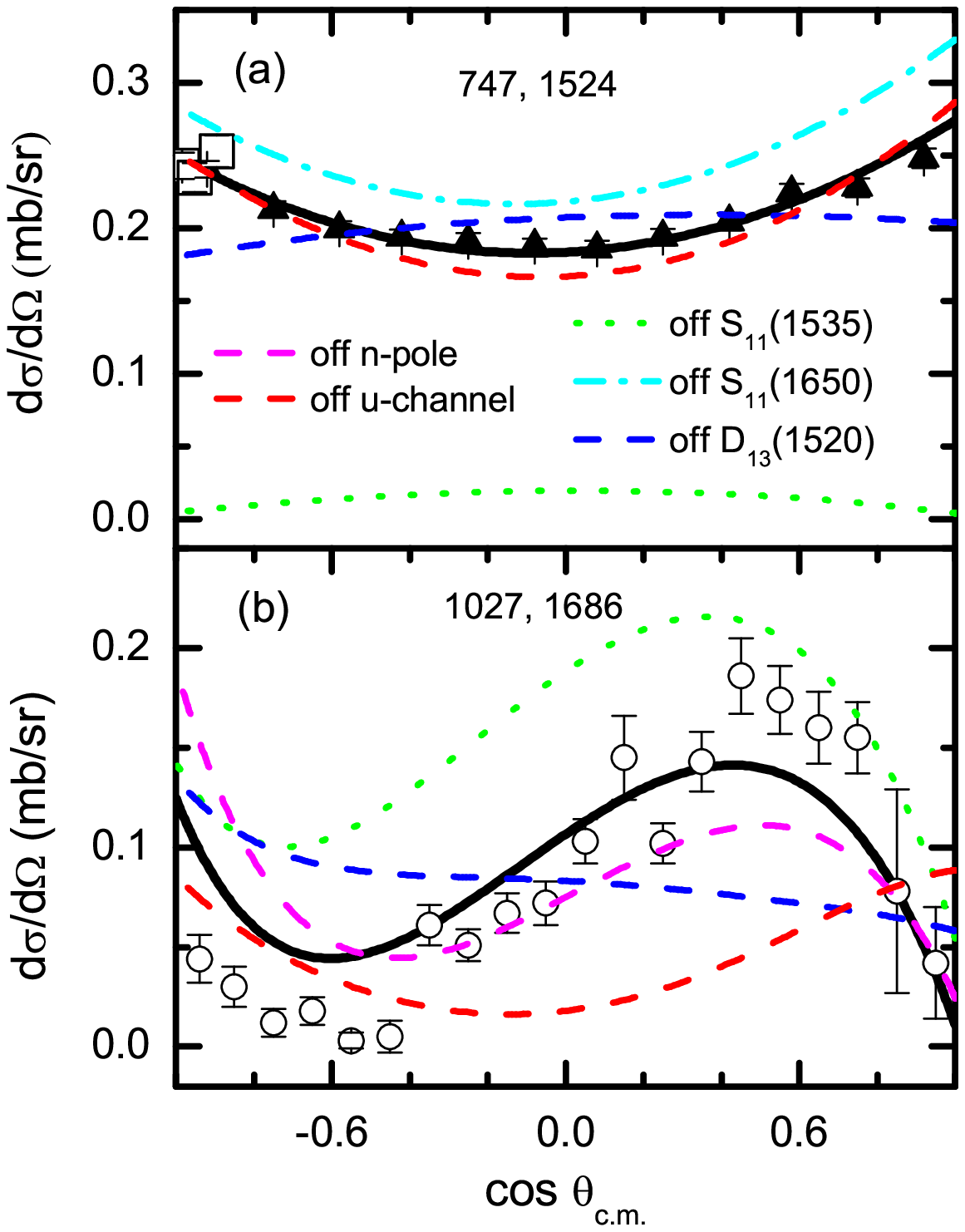}
\vspace{-0.2cm} \caption{ Differential cross sections of the
reaction $\pi^- p \to \eta n$ compared with experimental data at two
energy points $P_{\pi}$ =747, 1027 MeV/c. The bold solid curves
correspond to the full model result. The predictions by switching
off the contributions from $N(1535)S_{11}$, $N(1650)S_{11}$,
$N(1520)D_{13}$, and n-pole, $u$-channel backgrounds are indicated
explicitly by the legend in the figures.} \label{diff-etan-2}
\end{center}
\end{figure}

\subsection{$\pi^-p\rightarrow \eta n$}

The chiral quark model approach was first extended to the study of the reaction
$\pi^-p\rightarrow \eta n$ near threshold in our previous
work~\cite{Zhong:2007fx}. For the unthoroughness of partial wave
analysis for the $n=2$ shell resonances, and no considerations
of the configuration mixing between $N(1535)S_{11}$ and $N(1650)S_{11}$,
we only obtained preliminary results. In this work,
by combining the study of the reaction $\pi^-p\rightarrow K^0 \Lambda$, we present a
comprehensive study of the $\pi^-p\rightarrow \eta n$ process to
better understand this reaction and extract more reliable properties of nucleon resonances.

The differential cross sections and total cross section compared with
the experimental data are shown in Figs.~\ref{diff-etan} and~\ref{total-etan},
respectively. From the figures, it is seen that
the experimental data in the c.m. energy range from threshold up to $W\simeq 1.8$ GeV
can be reasonably described within the chiral quark model. Compared with
our previous study in~\cite{Zhong:2007fx}, the results in present work
have an obvious improvement.

In the $S$-wave states, dominant role of $N(1535)S_{11}$ can be found in the reaction. It is
responsible for the first hump around $P_{\pi}\simeq 0.76$ GeV/c ($W\simeq 1.5$ GeV). Without
the $N(1535)S_{11}$ contribution, the first hump disappears
completely. In addition, a sizeable contribution from $N(1650)S_{11}$
can be seen from Figs.~\ref{total-etan} and ~\ref{diff-etan-2}. Around
the first hump, $N(1650)S_{11}$ has obvious destructive
interferences with $N(1535)S_{11}$, which is consistent with our
previous study~\cite{Zhong:2007fx}. In the total cross section there
seems to exist another small bump structure around $P_{\pi}\simeq $
1.0 GeV/c ($W\simeq 1.7$ GeV). Based on our calculations shown in Fig.~\ref{total-etan},
the interferences between $N(1650)S_{11}$, $N(1535)S_{11}$ and
$u$-channel background might be responsible for the this structure,
which is different from our previous prediction~\cite{Zhong:2007fx}.
Our results in present work are consistent with the those analyses
within the coupled-channel approaches~\cite{Penner:2002ma,Shklyar:2004dy,Durand:2008kw,Durand:2008es}.

In the $D$-wave states, $N(1520)D_{13}$ plays an important role in the
reaction, which can be obviously seen in the differential cross
sections. Its interferences with the $N(1535)S_{11}$ and backgrounds
is crucial to produce the correct shape of the differential cross
sections in the whole energy region what we have considered. From
the Fig.~\ref{diff-etan} one can see that without the
$N(1520)D_{13}$ contribution, the shape of the differential cross
sections changes significantly. However, no obvious effects of
$N(1520)D_{13}$ on the total cross section can be found in the
$\pi^-p\rightarrow \eta n$ reaction. This feature was mentioned in
Refs.~\cite{Gasparyan:2003fp,Arndt:2005dg}. It should be pointed out
that to well describe the data, a large amplitude of $N(1520)D_{13}$
in the reaction is needed, which is about a factor of $2.18$ larger than that
derived in the SU(6)$\otimes$O(3) limit, which can not be explained
with configuration mixing effects.

With the energy increasing, the $u$-channel background become more and
more impportant in the reaction. Its large effects on both total cross
section and differential cross sections can be notably seen in the
energy region $P_{\pi}>$ 0.8 GeV/c ($W> 1.5$ GeV). Its interferences with the
resonances $N(1650)S_{11}$ and $N(1535)S_{11}$ are responsible for
the second bump structure around $P_{\pi}\simeq $ 1.0 GeV/c ($W\simeq 1.7$ GeV).

No determined evidence of the other resonances, such as
$D(1700)D_{13}$, $D(1675)D_{13}$, and higher $P$-, $F$-wave resonances is found
in the reaction. The background contributions from $n$-pole and
$t$-channel are less important to the reaction.

We should point out that some analyses of this reaction suggest
the need of the $N(1710)P_{11}$ resonance~\cite{Shklyar:2006xw,Penner:2002ma,Ronchen:2012eg,
Shklyar:2012js}. However, according to our analysis, no obvious
$N(1710)P_{11}$ contribution is required for a good description of the
experimental observations. Meanwhile, our analysis indicates that
$N(1720)P_{13}$ might have some effects on the cross sections. According to
our results from partial wave analysis (see Tab.~\ref{amplitudes}),
in the SU(6)$\otimes$O(3) symmetry limit the scattering amplitudes
of $N(1720)P_{13}$ are much larger than those of $N(1710)P_{11}$
around $W$=$1.7$ GeV. Thus, the role of $N(1720)P_{13}$ might be
more obvious than that of $N(1710)P_{11}$ if they are indeed seen in
the reaction, which is consistent with the chiral quark model study
in Ref.~\cite{He:2008uf}. In this work, we find that the role of
$N(1720)P_{13}$ is sensitive to its width. If we adopt a broad width
of $\Gamma\simeq 400$ MeV obtained by fitting the data, the contributions
of $N(1720)P_{13}$ to the reaction are negligibly small. However, if
we adopt a narrower width $\Gamma\simeq 120$ MeV as suggested in our
previous work by a study of the $\pi^0$ photoproduction~\cite{Xiao:2015gra}, we find
that the second bump structure in the total cross section become
more obvious, while around $W$=$1.7$ GeV the cross sections at
backward angles are enhanced significantly. It should be emphasized
that although our theoretical results seem to become bad compared
with the data with a narrow width of $N(1720)P_{13}$, we can not
exclude this possibility because the old data obtained many years
ago might be problematic for its uncontrollable uncertainties~\cite{Clajus:1992dh}.
New observations of the $\pi^-p\rightarrow \eta n$ reaction are urgently needed to better
understand the properties of nucleon resonances.

In brief, $N(1535)S_{11}$ play a dominant role in the $\pi^-p\to \eta n$
reaction near $\eta$ production threshold. In this low energy region,
$N(1650)S_{11}$ has notable destructive interferences with $N(1535)S_{11}$.
The $u$-channel background also plays a crucial role in the reaction. The
interferences between $N(1650)S_{11}$ and $u$-channel background might
be responsible for the second bump structure around $P_{\pi}\simeq$ 1.0 GeV/c ($W\simeq 1.7$ GeV).
$N(1520)D_{13}$ is crucial to describe the differential cross sections, although
it has small contributions to the total cross section.
To confirm the role of $N(1720)P_{13}$ in the reaction, new accurate measurements are urgently
needed in the c.m. energy range $W\simeq 1.5-1.8$ GeV. No obvious evidences of
$N(1700)D_{13}$, $N(1675)D_{15}$, $N(1710)P_{11}$,
and $N(1680)F_{15}$ are found in the $\pi^-p\to \eta n$ reaction, which is
in disagreement with the predictions in~\cite{Shklyar:2006xw,Penner:2002ma,Ronchen:2012eg,
Shklyar:2012js}, where the authors predicted the $N(1710)P_{11}$ resonance is needed to explain the reaction.

\section{Summary}

In this work, a combined study of the $\pi^-p\rightarrow K^0\Lambda$
and $\eta n$ have been carried out within a chiral quark model.
We have achieved reasonable descriptions of the data
in the c.m. energy range from threshold up to $W\simeq 1.8 $ GeV.

Obvious evidence of the $S$-wave nucleon resonances $N(1535)S_{11}$ and $N(1650)S_{11}$
is found in both of these two reactions. $N(1650)S_{11}$
contributes to the $\pi^-p\rightarrow K^0\Lambda$ reaction through configuration mixing with
$N(1535)S_{11}$. The determined mixing angle is $\theta_s\simeq 26.9^\circ$,
which is consistent with that extracted from $\eta$ and $\pi^0$
photoproduction processes in our previous works~\cite{Zhong:2011ti,Xiao:2015gra}.
Furthermore, the partial width ratios
$\Gamma_{\eta N}/\Gamma_{\pi N}$ and $\Gamma_{K\Lambda}/\Gamma_{\pi N}$ for these
$S$-wave states are extracted from the reactions, which
are close to the upper limit of the average values from the PDG
~\cite{Agashe:2014kda}. Obvious role of the $D$-wave state
$N(1520)D_{13}$ is found in $\pi^-p\rightarrow \eta n$ reaction, which has large effects on
the differential cross sections although its effects on the total cross section are tiny.
It should be pointed out that the effects of $N(1520)D_{13}$ on
the $\pi^-p\rightarrow K^0\Lambda$ are negligibly small.

The backgrounds play remarkable roles in these two strong interaction
processes. In the $\pi^-p\rightarrow K^0\Lambda$ process, the $u$-,
$t$-channel, and $n$-pole backgrounds have notable contributions to the cross sections.
While in the $\pi^-p\rightarrow \eta n$ process, the $u$-channel background plays
a crucial role in the higher energy region $W>1.5$ GeV.

The role of $P$-wave state $N(1720)P_{13}$ should be further confirmed
by future experiments. In present work, the data seem to favour a broad
width $\Gamma\simeq 400$ MeV for $N(1720)P_{13}$. However, our previous study of the $\pi^0$
photoproduction process indicates that the $N(1720)P_{13}$ might have a narrow
width of $\Gamma\simeq 120$ MeV~\cite{Xiao:2015gra}. If $N(1720)P_{13}$ has a broad
width of $\Gamma\simeq 400$ MeV, its contributions to the reactions
$\pi^-p\rightarrow K^0\Lambda$ and $\eta n$ should be negligibly small.
However, if $N(1720)P_{13}$ has a narrow width of $\Gamma\simeq 120$ MeV,
its contributions to the reactions are obvious, which can be seen from both
the total cross section and differential cross sections. The present data of
the $\pi^-p\rightarrow K^0\Lambda$ allow the appearance of
a narrow $N(1720)P_{13}$ resonance within the uncertainties. However,
when using a narrow width of $N(1720)P_{13}$ in the $\pi^-p\rightarrow \eta n$ reaction,
our theoretical results are notably larger than the
data at the backward angles. Improved measurements and more
reliable experimental data of the $\pi^-p\rightarrow K^0\Lambda$
and $\eta n$ reactions are needed to clarify the puzzle about $N(1720)P_{13}$.

Finally it should be pointed out that no obvious evidences of
$N(1700)D_{13}$, $N(1675)D_{15}$, $N(1710)P_{11}$,
and $N(1680)F_{15}$ are found in the $\pi^-p\rightarrow K^0\Lambda$
and $\eta n$ reactions, although they sit on the energy range
what we considered.

\section*{Acknowledgments}

This work is partly supported by the National Natural Science
Foundation of China (Grants No. 11075051 and No. 11375061), the
Hunan Provincial Natural Science Foundation (Grant No. 13JJ1018),
and the Hunan Provincial Innovation Foundation for Postgraduate.

\bibliographystyle{unsrt}

\end{document}